\documentclass[pdflatex,sn-mathphys-num]{sn-jnl}

\usepackage{graphicx}%
\usepackage{multirow}%
\usepackage{amsmath,amssymb,amsfonts}%
\usepackage{amsthm}%
\usepackage{mathrsfs}%
\usepackage[title]{appendix}%
\usepackage{xcolor}%
\usepackage{textcomp}%
\usepackage{manyfoot}%
\usepackage{booktabs}%
\usepackage{algorithm}%
\usepackage{algorithmicx}%
\usepackage{algpseudocode}%
\usepackage{listings}%

\theoremstyle{thmstyleone}%
\theoremstyle{thmstyletwo}%

\usepackage{amssymb,amsmath}
\usepackage{bm}
\usepackage{mathrsfs}

\usepackage{ulem}
\usepackage{xcolor}

\DeclareRobustCommand{\Erase}{\bgroup\markoverwith{\textcolor{red}{\rule[.5ex]{2pt}{2pt}}}\ULon}

\theoremstyle{thmstylethree}%

\begin{document}

\title[Numerical and experimental framework for bending elasticity of highly flexible slender structures]{Numerical and experimental framework for bending elasticity of highly flexible slender structures
}

\author[1]{\fnm{Shunsuke} \sur{Nomura}}

\author[1]{\fnm{Satsuki} \sur{Shibuya}}

\author[2]{\fnm{Isamu} \sur{Hashiguchi}\textsuperscript{\textdagger}}

\author[2]{\fnm{Ryuichi} \sur{Tarumi}}

\author*[1,3]{\fnm{Tomohiko G.} \sur{Sano}}\email{sano@mech.keio.ac.jp}

\affil[1]{\orgdiv{School of Integrated Design Engineering, Graduate School of Science and Technology}, \orgname{Keio University}, \orgaddress{\street{3-14-1 Hiyoshi,} \city{Yokohama}, \state{Kanagawa}, \postcode{2238522}, \country{Japan}}}

\affil[2]{\orgname{Graduate School of Engineering Science, Osaka University}, \orgaddress{\street{1-3 Machikaneyama}, \city{Toyonaka}, \state{Osaka}, \postcode{5608531}, \country{Japan}}}

\affil[3]{\orgdiv{Department of Mechanical Engineering, Faculty of Science and Technology}, \orgname{Keio University}, \orgaddress{\street{3-14-1 Hiyoshi}, \city{Yokohama}, \state{Kanagawa}, \postcode{2238522}, \country{Japan}}}


\abstract{
Slender structures are highly flexible, spanning several orders of magnitude in length scale. Their deformation depends on the slenderness of their cross sections, highlighting that the elasticity and geometry of structures are intrinsically coupled. The deformation of the cross-section becomes significant, particularly when tubes and pipes are subjected to bending, known as the Brazier instability. Although the bending performance of slender structures is quantified experimentally using a canonical three-point bending test, their numerical counterparts remain under-explored because complex contact mechanics must be implemented in simulations. In this study, we develop a computational framework to simulate experimental three-point bending tests using a hybrid material point method (hybrid-MPM) approach, which integrates Lagrangian finite element and Eulerian finite difference frameworks. We adapt our framework to elastic tubes and tape springs as canonical examples that exhibit characteristic bending deformation in which the cross-sectional and lengthwise bending are coupled. The predictions of numerical simulations are validated against desktop experiments and classical theory. The excellent agreement between the simulation and the experiments implies that the hybrid-MPM framework provides a robust computational framework for predicting the large deformation of structures involving complex contact, such as soft robots and deployable structures.
}


\maketitle
\renewcommand{\thefootnote}{\textdagger}
\footnotetext{Present address: Department of Micro Engineering, Graduate School of Engineering, Kyoto University, Kyoto, Japan.}

\section{Introduction}

{Thin structures bend and twist, exhibiting large deformations,} through their moment-curvature relations whose stiffness is characterized by their slenderness~\cite{Landau1980, Calladine1983, Vella2019, Holmes2019, Audoly2010, Bazant2010}. 
{The deformation of thin structures is inevitably coupled to the geometry of their slender cross sections.} The bending performance of slender structures, such as beams and pipes, is experimentally quantified using a canonical three-point bending test{, in which a structure spanning lengthwise two supports is pushed by a force sensor~\cite{ASTM-ThreePointB}.} 
Although this three-point bending test is widely used experimentally, its numerical counterpart has not been examined sufficiently. {The complexity of modeling the three-point bending test arises from contact interactions between the structure, supports, and indenter.}

\begin{figure}[h]
    \centering
    \includegraphics[width=\textwidth]{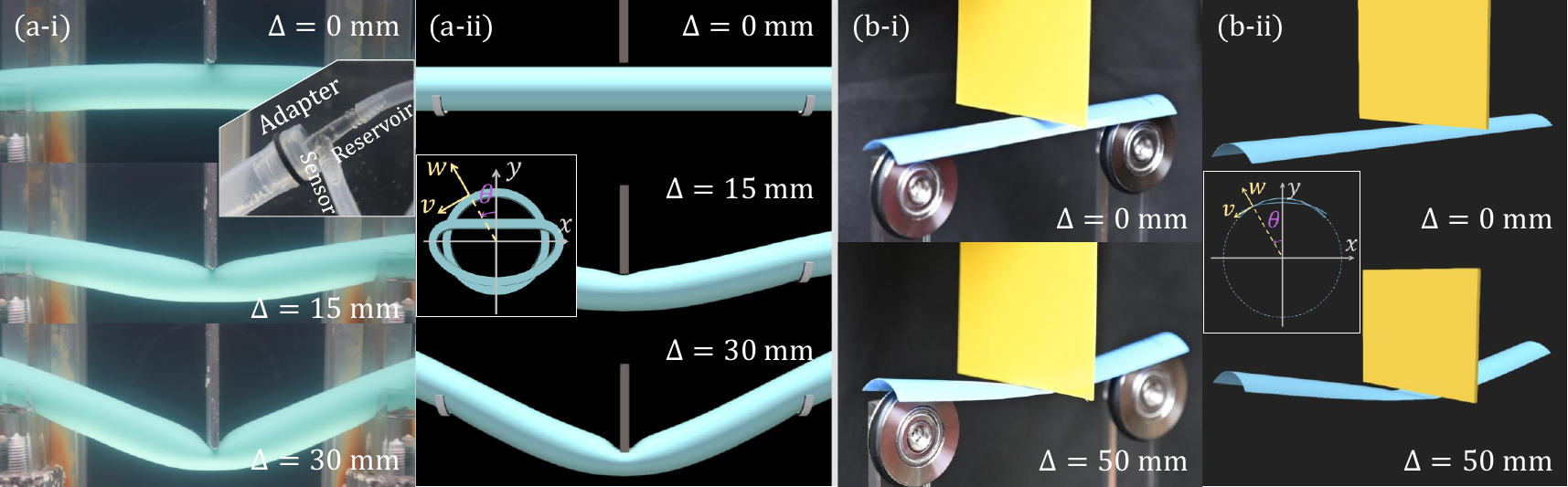}
    \caption{Bending test of hollow structures; (a) tubes and (b) tapes. Corresponding snapshots of our (a-i)(b-i) experiments and (a-ii)(b-ii) simulations.
    {(a-i) The deformation sequences of pressurized bending tubes of $(h,R) = (1.5,6.5)$~mm and $p = 2$~kPa. The inset shows a close-up view of the tube extremity, which is connected to the fluid reservoir and pressure sensor. (a-ii) Simulation snapshots of the tube under the same conditions as (a-i). 
    (b-i) The deformation process of tape springs with $(h,R,w) = (0.1,8.85,15.0)$~mm. The yellow rigid plate is moved downward to indent the tape spring. (b-ii) The corresponding simulation counterparts are shown. {The insets of (a-ii) and (b-ii) show the cross-sectional coordinate system and define the displacement components $v$ and $w$, together with the angle parameter $\theta$.}}
    }
    \label{fig:setup}
\end{figure}

In this paper, we focus on thin and hollow structures, i.e., tubes and tape springs (open cylindrical shells), as canonical examples that exhibit highly flexible deformations (Fig.~\ref{fig:setup}).
Tubes and tape springs span length scales from the micro to the kilometer scale. Hollow structures are lightweight and flexible, and are therefore used in both biological systems and engineering applications. Examples range from flagella and cilia, \textit{E.Coli}~\cite{Qiu2022}, plant tendrils and stems~\cite{Tani1961, Dumais2011, Bastien2015}, leaves~\cite{Grubb2008}, bendy straw~\cite{Bende2018}, insect wings~\cite{Saito2017}, carpenter's tape~\cite{Seffen1999,Guinot2012,Matsumoto2018}, tubes and pipes~\cite{Mahadevan2007}, soft actuators~\cite{Jones2021}, to structures such as pipelines~\cite{Gordon2009structures}. 
The characteristic deformation of hollow structures is the Brazier-like deformation, in which large cross-sectional deformation is coupled with lengthwise bending. {Apply an end-moment to a tube to form a uniform arc.} As the moment increases, the deformation becomes localized, followed by cross-sectional compression, as evidenced by kink formation~\cite{Brazier1927, Seide1961, Calladine1983, Ghatak2007, Qiu2022} (Fig.~\ref{fig:setup}(a)). 
The tape spring also exhibits similar kink formation as the familiar snapping sound of bent carpenter's tape~\cite{Seffen1999, Guinot2012, Seffen2019, Taffetani2019, Walker2018} (Fig.~\ref{fig:setup}(b)). 
Although Brazier-like instability is well studied theoretically and numerically, most studies are moment-controlled, examining moment-curvature relations. 
{
For example, Tanaka et al.~\cite{Tanaka2006} investigated the mechanical characteristics of buckling deformation in tubes under moment-controlled bending using finite element analysis based on shell theory.
In contrast, the mechanical response of slender structures is often evaluated experimentally using three-point bending tests, which involve large deformation and contact; these effects can be challenging to handle robustly with conventional FEM.
}
Characterization of force responses in realistic setups is crucial for leveraging the flexibility of tubes in engineering applications such as pneumatic or fluid-driven soft robots or deployable structures~\cite{Rus2015, Overvelde2015, Lee2017, Saito2017, Gorissen2020, El-Atab2020, Chi2022, Liu2022}. 

Here, we develop a computational model for three-point bending that combines the finite-element (FE) method and the material point method (MPM), using the Lagrangian and Eulerian frameworks alternately in the space-time integration~\cite{Sulsky1995, Jiang2015, Hu2019, deVaucorbeil2020, Qin2023}.
{
This hybrid-MPM framework combines the robustness of MPM in handling large deformation and complex contact with the accuracy of the FE description for structural deformation.
}
This combined method has been examined in simulations of soft robots, where the nonlinear deformation of structures under contact is relevant~\cite{Abe2025, Hashiguchi2026}.

\section{Review of bending performance of tubes and tape-springs}

We briefly review previous theoretical predictions of the bending performance of tubes and tape springs with length $\ell$, thickness $h$, and inner radius of the cross-section $R$, based on the classic literature~\cite {Brazier1927, Calladine1983}. We consider the three-point bending of tubes and tape springs, bridged at the supporting points separated by a distance $L$, where we measure the loading force from above, $F$, with the indenter displacement, $\Delta$. {We aim to derive the force–displacement relation by imposing a uniform curvature, $\kappa$, along the central axis of the tube.} 

The characteristic feature of tubes and tapes is the coupling between longitudinal bending and cross-sectional deformation. {As the tube bends, its cross-section ovalizes and flattens}, inducing a nonlinear bending response. The primary contributions of the nonlinear mechanical performance are the reduction of the moment of inertia, $I$, and the change of circumferential curvature, $K_{\theta}$. {Their mathematical definitions are provided separately for tubes and tape springs in Secs.~\ref{sec:TubeTheory} and~\ref{sec:TapeTheory}, respectively.} These values are calculated through the two-dimensional displacement of the cross-section, $\bm{u}(\theta) = w\hat{\bm{e}}_r + v\hat{\bm{e}}_{\theta}$ as a function of the angle from the $y$-axis in the natural configuration, $\theta$, where $(w,v)$ are radial and tangential components of the displacement (See Fig.~\ref{fig:setup}(a-ii)(b-ii) insets). For the remainder of this section, we set the longitudinal axis of tubes and tapes aligned with the $z$-axis and span the $x$-$y$ coordinate on the cross section whose origin matches the center of the arc. We assume the linear constitutive law with the Young's modulus, $E$, and the Poisson ratio, $\nu$. The indenter is lowered from above, $y>0$. 
In the following two subsections, we derive the moment-curvature relation for tubes (Sec.~\ref{sec:TubeTheory}) and tapes (Sec.~\ref{sec:TapeTheory}) based on this well-established idea, which will be compared with experiments and simulations later. {All the relevant parameters in our system are summarized in Table~\ref {tab:symbols}. }

\subsection{Bending elasticity of pressurized tube: Brazier instability}\label{sec:TubeTheory}

{We consider the bending deformation of the tube. The undeformed cross-section is described by $\bm{R}(\theta) = (-R\sin\theta, R\cos\theta)$, where $\theta$ is measured from the positive $y$-axis. The deformed configuration of the cross-section is denoted by $\bm{r}(\theta) = (x(\theta), y(\theta)) = \bm{R}(\theta) + \bm{u}(\theta)$. Upon bending, the cross-section is ovalized and becomes nearly elliptical, from which we assume the radial component of the displacement, $w$, as $w = - R\zeta \cos2\theta$. This assumption is based on Brazier's idea, where $\zeta$ is introduced as the normalized apex displacement. Its value is determined later by minimizing the total elastic energy. We further assume that the centerline of the cross-section is inextensible, $\epsilon_{\theta} = (dv/d\theta + w)/R = 0$, yielding the tangential component as $v = (R\zeta/2)\sin2\theta$.} The second moment of inertia in the configuration {with the deformed cross-section} is computed as $I = h\int_{0} ^{2\pi} y^2(\theta)d\theta = I_0 ^{\rm tube}\left(1 - 3\zeta/2 + O(\zeta^2)\right)$, where we define the second moment of inertia of the straight cylinder, $I_0 ^{\rm tube}\equiv\pi R^3h$, and drop the $O(\zeta^2)$ term as Brazier did. We note that the bending modulus is effectively reduced by changing the cross-section, as captured by the second term in $I$. 

We can now express the longitudinal bending energy of the tube per unit length, $U_{\ell}$, as $U_{\ell} = EI\kappa^2/\{2(1-\nu^2)\}${, which is realized by compressing and stretching the inner and outer arcs, respectively, of the tube cross section}. The change of circumferential curvature, $K_{\theta}$, is readily obtained as $K_{\theta}(\theta) = - (w + d^2w/d\theta^2)/R^2$, which gives the bending energy of the cross-section per unit length, $U_{\rm c}$, as $U_{\rm c} = (D/2)\int_0 ^{2\pi}K_{\theta}^2(\theta)Rd\theta$ with the bending rigidity, $D = Eh^3/\{12(1-\nu^2)\}$. Combining $U_{\ell}$ and $U_{\rm c}$, we get the total elastic energy per unit length as a function of $(\tilde{\kappa},\zeta)$, as
\begin{eqnarray}
    U(\tilde{\kappa},\zeta) = \frac{\pi Eh^3}{2(1-\nu^2)R}\left\{\left(1 - \frac{3}{2}\zeta\right)\tilde{\kappa}^2 + \frac{3}{4}\zeta^2\right\}\label{eq:Utube}. 
\end{eqnarray}
The relationship between $\tilde{\kappa} \equiv \kappa R^2/h$ and $\zeta$ is derived by minimizing Eq.~\eqref{eq:Utube} with respect to $\zeta$ as $\partial U/\partial\zeta = 0$, yielding, $\zeta = \tilde{\kappa}^2$. Finally, we reach the elastic energy per unit length as a function of imposed longitudinal curvature, $\kappa$, as
\begin{eqnarray}
    U(\tilde{\kappa}) = \frac{\pi Eh^3}{2(1-\nu^2)R}\left(\tilde{\kappa}^2 - \frac{3}{4}\tilde{\kappa}^4\right),
\end{eqnarray}
from which we obtain the moment-curvature relation as
\begin{eqnarray}
    M = \frac{\partial U}{\partial \kappa} = \frac{\pi EhR^3}{(1-\nu^2)}\left({\kappa} - \frac{3}{2}\frac{R^4}{h^2}{\kappa}^3\right)
\end{eqnarray}

To compare with the experimental force testing setup, we convert the moment curvature relation ($\kappa$ vs $M$) into the force displacement relation ($\Delta$ vs $F$). {The bending moment is generated by half of the applied force, $F/2$, acting with a lever arm of length $L/2$, giving $M=FL/4$.} {Assuming that the tube is sufficiently thin and undergoes uniform bending up to the Brazier instability, the indenter displacement is approximated as the deflection of the tube centerline. The displacement and curvature are therefore related by the geometric relation $\kappa = 8\Delta/L^2$}, which yields the force-displacement relation as
\begin{eqnarray}
    F = \frac{32\pi EhR^3}{(1-\nu^2)L^2}\left\{\frac{\Delta}{L} - 96\left(\frac{R^2}{hL}\right)^2\left(\frac{\Delta}{L}\right)^3\right\}.\label{eq:FD}
\end{eqnarray}
By introducing the dimensionless force and lengths as $\tilde{F} = F/F_0 ^{\rm tube}$ and {$\tilde{\Delta}\equiv\Delta/L$}, respectively, with the characteristic force and length scales, $F_0 ^{\rm tube}\equiv {32\pi EhR^3}/\{(1-\nu^2)L^2\}$ and $L$, we rewrite Eq.~\eqref{eq:FD} into the dimensionless form;
\begin{eqnarray}
    \tilde{F} = \tilde{\Delta} - 96\left(\frac{R^2}{hL}\right)^2\tilde{\Delta}^3,\label{eq:FD_res}
\end{eqnarray}
{For sufficiently small $\Delta$, the cubic term is negligible, and the force-displacement relation is approximately linear. The corresponding proportionality constant is defined as the initial bending stiffness, $K_0$. Applying the condition $dF/d\Delta=0$ to Eq. (5) gives the maximum force, $F_0^\ast$. Consequently, the initial stiffness, $K_0$, and the maximum force, $F_0^\ast$, are respectively derived as}
\begin{eqnarray}
    K_0 = \frac{32\pi EhR^3}{(1-\nu^2)L^3},~F^* _0=\frac{8\sqrt{2}\pi}{9}\frac{Eh^2R}{(1-\nu^2)L}.
\end{eqnarray}

In the presence of internal pressure $p$, the work done, accompanied by a change in cross-sectional shape, contributes to energy minimization. The pressure acts to prevent the deformation of the cross section, where the flexural hoop rigidity effectively increases by the factor of $(1 + p/p^*)$, where the characteristic pressure, $p^*\equiv Eh^3/4R^3$, indicates the effective stiffness associated with the elliptic bending mode~\cite{Calladine1983}. In other words, we multiply the bending energy by the above prefactor as
\begin{eqnarray}
    U(\tilde{\kappa},\zeta) = \frac{\pi Eh^3}{2(1-\nu^2)R}\left\{\left(1 - \frac{3}{2}\zeta\right)\tilde{\kappa}^2 + \frac{3}{4}(1+\tilde{p})\zeta^2\right\}\label{eq:Utube_p}, 
\end{eqnarray}
with the rescaled pressure, $\tilde{p} = p/p^*$. The relation between $\zeta$ and $\tilde{\kappa}$ is changed as $\zeta = (1+\tilde{p})^{1/2}\tilde{\kappa}^2$, from which we get the bending energy and the rescaled force-displacement relation as
\begin{eqnarray}
    U(\tilde{\kappa}) &=& \frac{\pi Eh^3}{2(1-\nu^2)R}\left(\tilde{\kappa} - \frac{3}{4}(1+\tilde{p})^{-1}\tilde{\kappa}^4\right)\\
    \tilde{F} &=& \tilde{\Delta} - \frac{96}{1+\tilde{p}}\left(\frac{R^2}{hL}\right)^2\tilde{\Delta}^3,\label{eq:FD_res_p}
\end{eqnarray}
respectively. The maximum force upon bending in the presence of internal pressure, $F^*$, is now given by
\begin{eqnarray}
    F^* = F_0 ^*\sqrt{1+ \frac{p}{p^*}}. \label{eq:Fstar_res_p}
\end{eqnarray}

\subsection{Bending elasticity of tape springs}\label{sec:TapeTheory}

We perform a similar analysis to the tape springs of width $2R\beta$, where the half of the opening angle is $\beta$, i.e., $\theta$ ranges as $-\beta\leq\theta\leq\beta$. {To construct the radial displacement of the cross-section, $w$, we expand $w$ with the Fourier series, leaving the long wavelength mode satisfying the moment-free boundary condition at $\theta = \pm\beta$. } Given that the circumferential moment is given by $M_{\theta} = DK_{\theta}$, the change of the curvature must vanish at $\theta = \pm\beta$, which yields the following functional form as
\begin{eqnarray}
    w = - R\zeta\cos\left(\frac{\pi}{2\beta}\theta\right).
\end{eqnarray}
Indeed, the change of the curvature, $K_{\theta} = (\zeta/R)(1-k^2)\cos(k\theta)$ with the wave number, $k\equiv\pi/(2\beta)$ satisfies the boundary conditions.
The tangential component is readily integrated as $v = - (R\zeta/k)\sin(k\theta)$. 
As we did for the Brazier instability, we compute the second moment of inertia in the deformed configuration up to $O(\zeta)$ as 
$I = \int_{-\beta} ^{\beta}(y(\theta)-\bar{y})^2 d\theta= R^3h(\tilde{I}_0(\beta) - 2\zeta\tilde{I}_1(\beta))$ 
with the second moment of inertia of the undeformed tape, $I_0 ^{\rm tape}=R^3h\tilde{I}_0(\beta)$, with the dimensionless parameters depending on the opening angle of the tape cross section,
\begin{eqnarray}
    \tilde{I}_0(\beta) &\equiv&\beta + \frac{\sin2\beta}{2} - \frac{2\sin^2\beta}{\beta}\simeq\frac{2}{45}\beta^5 + O(\beta^7)\label{eq:I0},\\
    \tilde{I}_1(\beta) &\equiv& \frac{2\beta}{\pi}\left(1 + \frac{k^2 + 2}{k^2-4}\cos2\beta - \frac{k^2 + 1}{k^2 -1}\frac{\sin2\beta}{\beta}\right)
    \simeq\left(2-\frac{\pi^2}{6}\right)\frac{8}{\pi^3}\beta^3 + O(\beta^5)\label{eq:I1}.
\end{eqnarray}
We note that the second moment of inertia, $I$, must be calculated about the centroid of the cross section as $(\bar{x}, \bar{y})$ with $\bar{x} = 0$ and $\bar{y}\equiv(1/\beta)\int_0^{\beta}y(\theta)d\theta$. 
{The right-hand-side approximations in Eqs. (\ref{eq:I0}) and (\ref{eq:I1}) are valid only in the shallow shell limit $(\beta \ll 1)$ and are presented for reference. Throughout this study, all theoretical predictions are obtained using the exact expressions in Eqs. (\ref{eq:I0}) and (\ref{eq:I1}).} 
Taking into account all the relevant physical parameters, we obtain the total elastic energy per unit length as a function of $(\tilde{\kappa},\zeta)$, as
\begin{eqnarray}
    U(\tilde{\kappa},\zeta) = \frac{Eh^3}{2(1-\nu^2)R}\left\{\left(\tilde{I}_0(\beta) - 2\zeta\tilde{I}_1(\beta)\right)\tilde{\kappa}^2 + \frac{\tilde{J}(\beta)}{12}\zeta^2\right\}\label{eq:Utape},
\end{eqnarray} 
with the dimensionless parameter appearing in the cross-sectional deformation, $\tilde{J} \equiv (1-(\pi/(2\beta))^2)^2\beta = (\pi/2)^4\beta^{-3} + O(\beta^{-1})$. 
The minimization of $U(\tilde{\kappa},\zeta)$ with respect to $\zeta$ allows us to eliminate $\zeta$ from Eq.~\eqref{eq:Utape}, i.e.,
\begin{eqnarray}
    U(\tilde{\kappa}) = \frac{Eh^3}{2(1-\nu^2)R} \left(\tilde{I}_0 (\beta)\tilde{\kappa}^2 - \frac{12\tilde{I}_1 (\beta)^2}{\tilde{J}(\beta)}\tilde{\kappa}^4\right),
\end{eqnarray}
from which the moment-curvature relation is readily obtained as
\begin{eqnarray}
    M = \frac{\partial U}{\partial \kappa} = \frac{EhR^3\tilde{I}_0}{1-\nu^2}\left({\kappa} -\frac{24\tilde{I}_1 ^2}{\tilde{I}_0\tilde{J}}\frac{R^4}{h^2}{\kappa}^3\right).
\end{eqnarray}
To compare with the experiments, we rewrite the above predictions in a dimensionless form as
\begin{eqnarray}
    \tilde{F} = \tilde{\Delta} - \frac{1536\tilde{I}_1 ^2}{\tilde{I}_0\tilde{J}} \left(\frac{R^2}{hL}\right)^2\tilde{\Delta}^3\label{eq:FD_tape_res},
\end{eqnarray}
where the force and displacement are normalized as $\tilde{F} = F/F_0 ^{\rm tape}$ with 
{$F_0 ^{\rm tape}\equiv {32EhR^3\tilde{I}_0}/\{(1-\nu^2)L^2\}$}
and $\tilde{\Delta} = \Delta/L$. The maximum force, $F^*$, is given by
\begin{eqnarray}
    F^* =\frac{2\sqrt{2}Eh^2R\tilde{I}_0^{3/2}\tilde{J}^{1/2}}{9(1-\nu^2)L\tilde{I}_1}.\label{eq:Fstar_tape}
\end{eqnarray}

\begin{table}[!h]
\centering
\caption{List of symbols used in three-point bending of tubes and tape springs.}
\label{tab:symbols}
\begin{tabular}{cl}
\hline
Symbol & Description \\
\hline
$D_{\mathrm{s}}$ & Diameter of the ball bearings \\
$E$ & Young's modulus \\
$F$ & Bending force \\
$\hat{F}$ & Maximum bending force obtained from experiments or simulations \\
$F^{*}$ & Theoretical maximum bending force \\
$h$ & Thickness of the tubes or the tape springs \\
$L$ & Distance between the supports \\
$p$ & Internal pressure \\
$\tilde{p}$ & Dimensionless internal pressure \\
$p^*$ & Characteristic internal pressure \\
$R$ & Inner radius of the tubes or the tape springs \\
$R_{\mathrm{m1}}$ & Radius of the acrylic tube for mold \\
$R_{\mathrm{m2}}$ & Radius of stainless steel cylinder for mold \\
$w$ & Tape spring width (along the arc)\\
$\Delta$ & Indenter displacement \\
$\beta$ & Half of the subtended angle of the tape springs\\
$\nu$ & Poisson's ratio \\
\hline
\end{tabular}
\end{table}

\section{General framework of Material Point Method}\label{sec:sim_framework}

{In this section, we formulate the computational framework to simulate the bending elasticity of tubes and tapes by combining FEM and MPM. Our formulation is based on Refs.~\cite{Sulsky1995,Jiang2015, Hu2019, deVaucorbeil2020, Hinton1976, Wu2006, Hu2018} which has been validated experimentally~\cite{Abe2025, Hashiguchi2026}. 
In Sec.~\ref{sec:FEM}, we formulate the general formalism of FEM based on the equation of motion within the weak form. In Sec.~\ref{sec:MPM}, we detail the MPM calculation protocols to take into account the contact mechanics and complete the hybrid-MPM formalism employed in this study. 

{Numerical approaches for the analysis of highly flexible slender structures have been extensively developed, with the finite element method (FEM) being the most widely used framework because of its high accuracy in structural mechanics~\cite{Zienkiewicz2013}. However, large deformation accompanied by self-contact or contact with multiple bodies often requires sophisticated contact algorithms and remeshing strategies, which may reduce computational robustness~\cite{Sano2022, Baek2021}. The Material Point Method (MPM) has recently emerged as an alternative approach for large-deformation problems owing to its hybrid Lagrangian–Eulerian description, which naturally handles severe deformation and contact without mesh distortion~\cite{Jiang2016}. The hybrid-MPM framework adopted in the present study combines the accuracy of FEM in describing shell elasticity with the robustness of MPM in contact mechanics. While the computational cost is generally higher than that of conventional FEM, this framework provides a unified approach for simulating highly flexible structures undergoing large deformation and complex contact, as demonstrated recently~\cite{Abe2025}.}
}

{For completeness of the literature reviews for computational approaches of highly flexible slender structures, we note the reduced simulation framework for their deformation, such as the centerline-based approach for rod and ribbon-like structures, whose deformations are represented by those of the centerline with the cross-sectional deformation being neglected. This method is widely used to simulate filamentary structures at different length scales, such as DNA, helical bacteria, and human hair, for which cross-sectional deformation is negligible~\cite{Chirico1994, Bergou2008, Sano2022sim}, and it has also been developed in the computer graphics community~\cite{BACQLL06, Bergou2008}.
This approach is known to be efficient for simulating the real-time dynamics of slender structures~\cite{Bergou2008}, whereas it is not applicable when cross-sectional deformation is relevant. Although this reduced approach has been extended to shell structures using the differential geometry approach~\cite{Huang2025} and frictional contact models~\cite{Sano2023, Crassous2023}, we adopt the hybrid-MPM framework detailed below to simulate the deformation of highly flexible, slender structures with relevant cross-sectional deformation under realistic experimental conditions.
}

\subsection{Finite element discretization for equation of motion}\label{sec:FEM}

We start with the equation of motion (EoM) of the material and explicitly state the boundary condition. The equation of motion (strong form) is rewritten into a weak form by introducing the test function, $\bm{w}$. Based on the weak form formulation, we derive the discretized version of the EoM.

Within the total Lagrangian framework of continuum mechanics, the motion of a body is described with respect to the reference configuration.
Let $\Omega_0 \subset \mathbb{R}^3$ and $\Omega \subset \mathbb{R}^3$ denote the domains occupied by the body in the reference and current configurations, respectively.
The motion is described by the mapping $\bm{\varphi} : \Omega_0 \to \Omega$, which maps a material point $\bm{X}\in\Omega_0$ in the reference configuration to its current position $\bm{x}\in\Omega$, and the displacement field is accordingly defined by $\bm{u}=\bm{x}-\bm{X}$.
The boundary $\Gamma_0$ of the domain $\Omega_0$, {whose (outer) surface normal is denoted by $\bm{N}$}, is divided into the Dirichlet boundary $\Gamma_D$ and the Neumann boundary $\Gamma_N$, which satisfy $\Gamma_D \cup \Gamma_N = \Gamma_0$ and $\Gamma_D \cap \Gamma_N = \emptyset$.
In conventional FEM contact simulations, the strong form includes explicit contact conditions governing interactions between deformable bodies. In the present hybrid–MPM simulation, however, the contact interaction is not treated explicitly, but is instead handled through the MPM mapping operation described in the next section.
The governing equations in the domain $\Omega_0$ are expressed as follows:
\begin{align}
\label{eq : strong}
    \rho\ddot{\bm{x}} &= \text{Div}\textbf{P} + \bar{\bm{b}} \quad&&\text{in}\quad \Omega_0, \\
\label{eq : Neumann}
    \textbf{P}\bm{N} &= \bar{\bm{t}} \quad&&\text{on}\quad \Gamma_N, \\
\label{eq : Dirichlet}
    \bm{x} &= \bar{\bm{x}}\quad &&\text{on}\quad\Gamma_D,
\end{align}
where $\rho$ is the mass density, $\bar{\bm{b}}$, $\bar{\bm{t}}$, and $\bar{\bm{x}}$ denote the body force in $\Omega_0$, traction on $\Gamma_N$, and position on $\Gamma_D$, respectively.
{Equation~\eqref{eq : strong} represents the equations of motion (strong form) of a material point, $\bm{x}$, whose Neumann and Dirichlet boundary conditions are respectively given by Eqs~\eqref{eq : Neumann} and \eqref{eq : Dirichlet}. }
$\ddot{\bm{x}} = \partial^2 \bm{x} / \partial t^2$ denotes the acceleration field. Hereafter, the superposed dot $\dot{(\,)}$ denotes the time derivative.
$\mathrm{Div}(\cdot)$ represents the divergence operator taken with respect to the reference coordinate $\bm{X}$.
{$\textbf{P} = \partial W / \partial \textbf{F}$} denotes the first Piola--Kirchhoff stress tensor, which is obtained by differentiating the strain energy density function {$W$} with respect to the deformation gradient tensor $\textbf{F} = \partial \bm{x} / \partial \bm{X}$.
The form of the function {$W$} depends on the constitutive law of the elastic body, and in the present study, the Neo-Hookean model is employed as follows, 
{
\begin{align}
    W = \frac{\mu}{2}\left(I_1-3\right) - \mu\text{ln}(J) + \frac{\lambda}{2}\text{ln}^2(J),
\end{align}
}
where $I_1=\text{tr}(\textbf{F}^T\textbf{F})$ and $J = \text{det}(\textbf{F})$ denote the first invariant of the right Cauchy--Green deformation tensor and the Jacobian determinant, respectively.
$\mu$ and $\lambda$ are the Lam\'e parameters, which are expressed in terms of the material parameters, Young's modulus $E$ and Poisson's ratio $\nu$, as $\mu = {E}/\{2(1 + \nu)\}$ and {$\lambda = {E\nu}/\{(1+\nu)(1-2\nu)\}$}.

To derive the weak form of the governing equation, we take the inner product of the strong form equation~(\ref{eq : strong}) with a test function $\bm{w}$ and integrate it over the domain $\Omega_0$ as follows:
\begin{align}
\label{eq : weak}
    \int_{\Omega_0}\rho\ddot{\bm{x}}\cdot{\bm{w}}dV = \int_{\Omega_0}\text{Div}\textbf{P}\cdot\bm{w}dV + \int_{\Omega_0}\bar{\bm{b}}\cdot\bm{w}dV.
\end{align}
Applying the divergence theorem to the first term on the right-hand side of Eq.~(\ref{eq : weak}) and substituting Eq.~(\ref{eq : Neumann}), the following weak form equation is obtained:
\begin{align}
\label{eq : weak form of the EOM}
    \int_{\Omega_0}\rho\ddot{\bm{x}}\cdot{\bm{w}}dV = -\int_{\Omega_0}\textbf{P}:\frac{\partial\bm{w}}{\partial\bm{X}}dV + \int_{\Gamma_N}\bar{\bm{t}}\cdot\bm{w}d\Gamma + \int_{\Omega_0}\bar{\bm{b}}\cdot\bm{w}dV.
\end{align}
Note that the condition $\bm{w}=\bm{0}$ on the boundary $\Gamma_D$ is used in deriving the above equation.

In the FEM simulation, the spatial coordinate $\bm{x}$ and the test function $\bm{w}$ are approximated using nodal values and shape functions as follows:
\begin{align}
\label{eq : approx}
    \bm{x}\approx\sum_{\alpha\in\mathcal{F}}K_\alpha\bm{x}_\alpha, \quad \bm{w} \approx\sum_{\alpha\in\mathcal{F}}K_\alpha\bm{w}_\alpha,
\end{align}
where $\mathcal{F}$ and $K_{\alpha}$ denote the set of all finite element nodes {and shape function, respectively}.
In this study, 27-node second-order Lagrange hexahedral elements~\cite{Komatitsch1999} are employed, and the corresponding quadratic shape functions, $K_{\alpha}$, are used for the spatial discretization.
The subscript $\alpha$ indicates that the quantity is associated with node $\alpha$. 
Note that the Einstein summation convention is not adopted in this study. 
Summation is performed only when the summation symbol is explicitly shown, as in Eq. (\ref{eq : approx}).
Substituting the approximated expressions~(\ref{eq : approx}) into the weak form of the equation of motion~(\ref{eq : weak form of the EOM}) and applying the row-sum lumped mass formulation~\cite{Hinton1976, Wu2006}, we obtain the following discretized equation of motion at node $\alpha$:
\begin{align}
\label{eq : discretized EOS}
    m_\alpha\dot{\bm{v}}_\alpha = \bm{f}_\alpha^{\text{int}} + \bm{f}_\alpha^{\text{ext}}\quad (\alpha\in\mathcal{F}),
\end{align}
where $m_\alpha$, $\bm{f}_\alpha^{\mathrm{int}}$, and $\bm{f}_\alpha^{\mathrm{ext}}$ are the lumped nodal mass, internal force, and external force associated with node $\alpha$, respectively. 
{Note that we do not take the summation over the index $\alpha$ on the left-hand side of \eqref{eq : discretized EOS}.}
These quantities are given by
\begin{align}
\label{eq : lumped mass}
    m_\alpha &= \int_{\Omega_0}\rho K_\alpha dV, \\
    \bm{f}_\alpha^\text{int} &= - \int_{\Omega_0}\textbf{P}\frac{\partial K_\alpha}{\partial\bm{X}} dV, \\
    \bm{f}_\alpha^\text{ext} &= \int_{\Gamma_N}\bar{\bm{t}}K_\alpha d\Gamma + \int_{\Omega_0}\bar{\bm{b}}K_\alpha dV.
\end{align}
By introducing the mass-proportional damping term, as commonly used in the dynamic relaxation method~\cite{OAKLEY199567}, so that the system relaxes to mechanical equilibrium within a reasonable computational time, we obtain the discretized EoM as
\begin{align}
\label{eq : discretized EOS with dumping}
    m_\alpha\dot{\bm{v}}_\alpha = -\gamma_{\rm d} m_{\alpha}\bm{v}_\alpha + \bm{f}_\alpha^{\text{int}} + \bm{f}_\alpha^{\text{ext}}\quad (\alpha\in\mathcal{F}),
\end{align}
where $\gamma_{\rm d} > 0$ is the damping coefficient per unit mass. 
By applying the forward Euler time discretization to the acceleration term on the left-hand side of Eq.~\eqref{eq : discretized EOS with dumping} with the time increment $\Delta t$, $\dot{\bm{v}}_\alpha = (\bm{v}_\alpha^\ast-\bm{v}_\alpha^{(n)}) / \Delta t$, an independent velocity update equation is obtained for each node as follows:
{
\begin{align}
\label{eq : updated velocity by FEM}
    \bm{v}_\alpha^\ast = \frac{2-\Delta t \gamma_{\rm d}}{2+\Delta t \gamma_{\rm d}}\bm{v}_\alpha^{(n)} + 
    \frac{2\Delta t}{m_\alpha(2 + \Delta t\gamma_{\rm d})}\left(\bm{f}_\alpha^{\text{int(n)}} + \bm{f}_\alpha^{\text{ext(n)}}\right),
\end{align}
}
where the superscript $(n)$ indicates that the variable is associated with time $t = n \Delta t$. $\bm{v}_\alpha^\ast$ denotes the {temporal} updated velocity obtained from $\bm{v}_\alpha^{(n)}$ through the explicit FEM simulation. {This temporal velocity $\bm{v}_\alpha^\ast$ is used for MPM analysis when we implement contact mechanics as detailed next. }

\subsection{Mapping operation of MPM for contact analysis}\label{sec:MPM}
The velocity update equation given in Eq.~(\ref{eq : updated velocity by FEM}) does not include the contact force on the right-hand side and therefore does not account for interactions with other deformable bodies.
In this section, we present a procedure for obtaining the contact-corrected velocity $\bm{v}_\alpha^{(n+1)}$ {in the next time step $n+1$} from the intermediate velocity $\bm{v}_\alpha^\ast$ by applying the MPM mapping operation.

\begin{figure}[h]
    \centering
    \includegraphics[width=\textwidth]{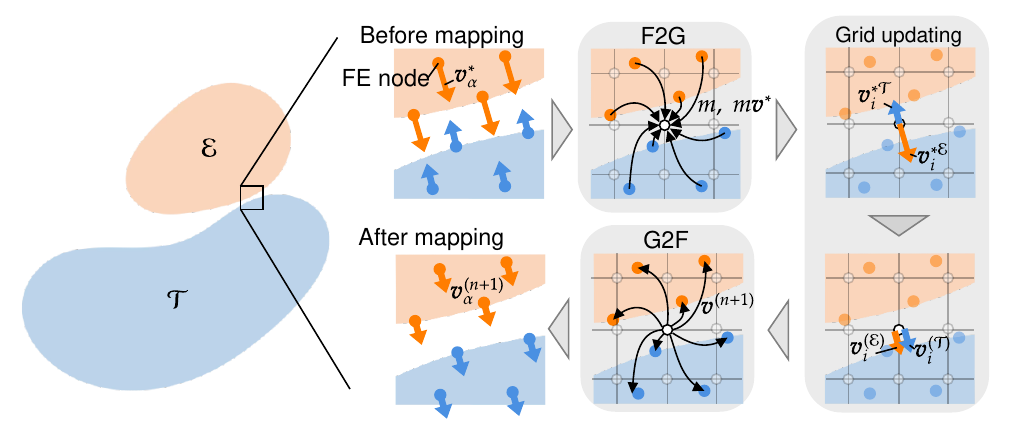}
    \caption{Schematic illustration of the mapping operation used to obtain the contact-corrected velocity $\bm{v}_\alpha^{(n+1)}$ from the intermediate FE nodal velocity $\bm{v}_\alpha^\ast$ by incorporating contact interactions based on the contact method of Bardenhagen et al.~\cite{BARDENHAGEN2000529}}
    \label{fig:mpm}
\end{figure}

Owing to the characteristics of the {conventional} MPM algorithm, the non-penetration condition between multiple bodies is automatically satisfied. 
{However, this approach may cause non-physical sticking by artificially constraining tangential slip at the contact boundary~\cite{BARDENHAGEN2000529}.}
To overcome these limitations, the contact algorithm proposed by Bardenhagen et al.~\cite{BARDENHAGEN2000529} is combined with dynamic FEM simulation in this study.
The mapping operation of this method, {which is schematically shown in Fig.~\ref{fig:mpm},} consists of three steps:
(i) the FE node-to-grid (F2G) step, in which the velocities of the FE nodes are mapped onto the Eulerian grid nodes (Sec.~\ref{sec:F2G});
(ii) the grid updating step, in which contact detection and velocity correction are performed on the Eulerian grid (Sec.~\ref{sec:GridUp}); and
(iii) the Grid-to-FE node (G2F) step, in which the updated velocities on the Eulerian grid are mapped back to the FE nodes (Sec.~\ref{sec:G2F}).
The details of each step are summarized below.

\subsubsection{Finite Element node-to-grid (F2G) step}\label{sec:F2G}

{The physical quantities calculated so far are defined on the Lagrangian (FE) node, which will be mapped onto the Eulerian node. We use roman indices, $i,j,\cdots$, for the Eulerian node instead of Greek indices, $\alpha,\beta,\cdots$, used so far, to distinguish them. The mapping calculation is employed with the weight function $\psi$. }
The quantities at the Eulerian grid nodes at $i$ are computed from the corresponding FE nodal values as follows:
\begin{align}
\label{eq : f2g}
    m_i^{\mathcal{Q}} = \sum_{\alpha\in\mathcal{F}_\mathcal{Q}}\psi(\bm{d}_{\alpha i}^{(n)}) m_\alpha,\quad
    (m\bm{v}^\ast)_i^{\mathcal{Q}} = \sum_{\alpha\in\mathcal{F}_\mathcal{Q}}\psi(\bm{d}_{\alpha i}^{(n)}) m_\alpha\left(\bm{v}_\alpha^\ast - \bm{C}_\alpha^{(n)}\bm{d}_{\alpha i}^{(n)}\right),
\end{align}
where $m_i$ and $(m\bm{v}^\ast)_i$ denote the mass and momentum at the Eulerian grid node $i$, respectively.
The superscript $\mathcal{Q} \in \{\mathcal{T}, \mathcal{E}\}$ is used to distinguish variables for different bodies. In the following, $\mathcal{T}$ denotes variables associated with the material of interest, such as the tube and the tape, while $\mathcal{E}$ denotes variables associated with external bodies, such as the indenter and support.
$\mathcal{F}_{\mathcal{Q}}$ denotes the set of all FE nodes belonging to body $\mathcal{Q}$.
$\bm{C}_\alpha$ denotes the affine velocity tensor at FE node $\alpha$, a second-order tensor used to preserve the velocity gradient lost during the mapping process. For further details, see Refs.~\cite{Jiang2015, Hu2018}.
$\psi$ denotes the weighting function that depends on the relative position $\bm{d}_{\alpha i}^{(n)} = \bm{x}_\alpha^{(n)} - \bm{x}_i$ of FE node $\alpha$ with respect to Eulerian grid node $i$. 
In this study, the weighting function $\psi$ is expressed in tensor-product form using the basis function $\psi_B$ in each spatial direction, i.e., $\psi(\bm{d}_{\alpha i})=\psi_B(d_x)\psi_B(d_y)\psi_B(d_z),$ where $d_j$ $(j \in \{x, y, z\})$ denotes the $j$-th component of the vector $\bm{d}_{\alpha i}$, and $\psi_B$ is the quadratic B-spline basis function~{\cite{Steffen2008}} in each spatial direction.
The function $\psi_B$ is given as follows:
\begin{align}
    \psi_B(d) = \begin{cases}
        \dfrac{1}{2}\left(\dfrac{d}{\Delta g} + \dfrac{3}{2}\right)^2 & \left(-\dfrac{3}{2}\leq\dfrac{d}{\Delta g}\leq-\dfrac{1}{2}\right) \\[6pt]
        -\left(\dfrac{d}{\Delta g}\right)^2 + \dfrac{3}{4} & \left(-\dfrac{1}{2}\leq\dfrac{d}{\Delta g}\leq\dfrac{1}{2}\right) \\[6pt]
        \dfrac{1}{2}\left(\dfrac{d}{\Delta g} - \dfrac{3}{2}\right)^2 & \left(\dfrac{1}{2}\leq\dfrac{d}{\Delta g}\leq\dfrac{3}{2}\right) \\[6pt]
        0 & \left(\text{otherwise}\right),
    \end{cases}
\end{align}
where {$\Delta g$} denotes the grid spacing. {Although a uniform grid spacing $\Delta g$ is employed in this study, the formulation can be extended to non-uniform grid spacings in the $x$-, $y$-, and $z$-directions.}

\subsubsection{Grid updating step}\label{sec:GridUp}

{We have distributed mass and momentum onto the Eulerian grid in Eq.~\eqref{eq : f2g}, which allows us to calculate the contact mechanics between different bodies.}
From the mass and momentum given in Eq.~(\ref{eq : f2g}), the intermediate velocity at each Eulerian grid node, $\bm{v}_i^{\ast\mathcal{Q}} \equiv (m\bm{v}^\ast)_i^\mathcal{Q} / m_i^\mathcal{Q}$, which does not include contact interaction, is obtained.
{By taking into account the contact mechanics, we update the intermediate velocity, $\bm{v}_i^{\ast\mathcal{Q}}$, as the corrected one $\bm{v}_i^{\mathcal{Q}}$, which will be used in Sec.~\ref{sec:G2F} to compute the velocity in the next time step.}

In the contact algorithm of Bardenhagen et al.~\cite{BARDENHAGEN2000529}, contact is detected based on the normal component of the relative intermediate velocity between the bodies, $(\bm{v}_i^{\ast\mathcal{T}} - \bm{v}_i^{\ast\mathcal{E}})\cdot\bm{n}_i$, and the contact-corrected velocity $\bm{v}_i^\mathcal{Q}$ is subsequently computed. {Here, the surface normal vector at the contact surface of body $\mathcal{E}$} is denoted as $\bm{n}_i$. 
When two structures move to separate from each other, i.e., when $(\bm{v}_i^{\ast\mathcal{T}} - \bm{v}_i^{\ast\mathcal{E}})\cdot\bm{n}_i > 0$, contact interaction does not need to be taken into account (non-adhesive). Therefore, the corrected velocity is set to be identical to the intermediate velocity as $\bm{v}_i^\mathcal{Q} = \bm{v}_i^{\ast\mathcal{Q}}$.
On the other hand, when they are approaching as $(\bm{v}_i^{\ast\mathcal{T}} - \bm{v}_i^{\ast\mathcal{E}})\cdot\bm{n}_i \leq 0$, $\bm{v}_i^{\mathcal{Q}}$ is obtained {by satisfying the momentum conservation} as follows:
\begin{align}
    \bm{v}_i^{\mathcal{T}} = \frac{1}{m_i^{\mathcal{T}}}\left( (m\bm{v}^\ast)_i^{\mathcal{T}} + \Delta t\,\bm{f}_i \right),\quad \bm{v}_i^{\mathcal{E}} = \frac{1}{m_i^{\mathcal{E}}}\left( (m\bm{v}^\ast)_i^{\mathcal{E}} - \Delta t\,\bm{f}_i \right), \label{eq:vTE}
\end{align}
where $\bm{f}_i$ denotes the normal contact force exerted on body $\mathcal{T}$ by body $\mathcal{E}$. {Hereafter, we neglect the tangential (friction) forces at contact, which is sufficient to reproduce our experimental results.} 
By enforcing the non-penetration condition, $(\bm{v}_i^{\mathcal{T}} - \bm{v}_i^{\mathcal{E}})\cdot\bm{n}_i = 0${, with the aid of Eq.~\eqref{eq:vTE}}, the (normal) contact force $\bm{f}_i$ is given by
\begin{align}
\label{eq : f_i}
    \bm{f}_i =  f_n\bm{n}_i\quad\text{with}\quad f_n = \frac
    {m_i^\mathcal{T}(m\bm{v}^\ast)_i^\mathcal{E} - m_i^\mathcal{E}(m\bm{v}^\ast)_i^\mathcal{T}}
    {\Delta t\left(m_i^\mathcal{T} + m_i^\mathcal{E}\right)}\cdot\bm{n}_i.
\end{align}

\subsubsection{Grid-to-Finite Element node (G2F) step}\label{sec:G2F}

{We have implemented the contact condition between two bodies, whose Eulerian velocity field is updated as $\bm{v}_i^\mathcal{Q}$ (See Eq.~\eqref{eq:vTE}). We now transform the Eulerian velocity field into the nodal velocities, $\bm{v}_\alpha$, by ``inverting" the mapping protocol in Sec.~\ref{sec:F2G}.}
The nodal velocity $\bm{v}_\alpha$ and the affine velocity tensor $\bm{C}_\alpha$ for each body are obtained as follows{~\cite{Jiang2015, Hu2018}}:
\begin{align}
    \bm{v}_\alpha^{\mathcal{Q}(n+1)} =& \sum_{i\in\mathcal{G}}\psi(\bm{d}_{\alpha i})\bm{v}_i^\mathcal{Q}, \\
    \bm{C}_\alpha^{\mathcal{Q}(n+1)} =& \frac{4}{\Delta g}\sum_{i\in\mathcal{G}}\psi(\bm{d}_{\alpha i})\bm{v}_i^\mathcal{Q}\otimes(- \bm{d}_{\alpha i}),
\end{align}
where $\mathcal{G}$ denotes the set of all Eulerian grid nodes.
The velocities at the nodes on the Dirichlet boundary are replaced with the prescribed values (e.g., velocity is set to be zero for clamped boundary conditions).
Finally, the nodal positions are advanced using the updated velocity.
\begin{align}
    \bm{x}_\alpha^{\mathcal{Q}(n+1)} = \bm{x}_\alpha^{\mathcal{Q}(n)} + \Delta t\bm{v}_\alpha^{\mathcal{Q}(n+1)}.
\end{align}
By repeating these protocols, we simulate the dynamics of structures up to the necessary time step. 

We have introduced the general framework of hybrid-MPM in this section. In the following two sections, we apply this framework to the canonical mechanical testing of two different highly flexible structures:~{Sec.~\ref {sec:Tube}, the pressurized tube, and Sec.~\ref {sec:Tape} the tape spring.}

\section{Example 1: Pressurized cylindrical tubes}\label{sec:Tube}

We now simulate the three-point bending test to quantify the bending performance of the pressurized tube. We summarize the simulation and experimental setup in Secs.~\ref{sec:TubeSim} and ~\ref{sec:TubeExp}, respectively. Their results are summarized and validated in Sec.~\ref{sec:TubeComparison}.

\subsection{Simulation setting}\label{sec:TubeSim}

The three-point bending simulation of the elastomeric tube is performed using the hybrid-MPM framework presented in Section~\ref{sec:sim_framework}.
The geometries of each component are defined as follows:
The tube is a cylindrical shell with an inner radius of $R = 6.5$~mm, a thickness $h$, and a length of 200~mm. The Young modulus and Poisson's ratio are $E = 0.47$~MPa and $\nu = 0.5$, respectively.
{We set the tube thickness as $h = 0.9$~mm, $1.2$~mm or $1.5$~mm. The simulation results for $h = 1.2$~mm and $ 1.5$~mm are used to compare with experiments, whereas those for $h = 0.9$~mm clarify the systematic effect of tube thickness.}
The tube extremities are capped by the same material of 1~mm thickness to ensure that the tube is pressurized from inside.
The indenter is a narrow rectangular plate with a height of 30~mm, a width of 40~mm, and a thickness of 3~mm. The Young's modulus is set to be $E=3.2$~GPa so that the deformation of the indenter is negligible (almost rigid body).
The supports are modeled as {rigid thin arcs} with an inner radius of 10~mm, a thickness of 1~mm, and a width of 3~mm, {thereby preventing transverse translation of the tube. The axes of the supports align with the axis of the tube}.
The indenter is placed in the middle of the supports and the tube, consistent with experiments. 
The distance between the supports is $L = 130$~mm.

An external force $\bm{f}_{\alpha}^{\mathrm{pres}}$, proportional to the internal pressure~{$p$}, is applied to the inner surface of the tube, which is included in the equation of motion to $\bm{f}_{\alpha} ^{\rm ext}$ in Eq.~\eqref{eq : discretized EOS}. 
The force acting on each node on the inner surface of the tube is given by the integral of the pressure over the inner surface of the tube, {$\Gamma_N ^{\rm in}$}, as {$\bm{f}_\alpha^\text{pres} = -\int_{\Gamma_N ^{\rm in}}p\bm{n}K_\alpha d\Gamma$}. 
Here, the inner surface of the tube (including caps) is denoted by $\Gamma_N ^{\rm in}$,
$\bm{n}$ is the outward unit normal vector to the surface {in the current configuration}, and $K_\alpha$ is the shape function associated with node $\alpha$.
The simulation and measurement protocol is as follows. The tube is pressurized from the inside by {$p$} without indentation, and the pressure is held constant long enough for the pressurized tube to remain straight before indentation. {As shown in the insets of Fig. 3, the pressurized tube expands as the internal pressure increases. Although the internal pressure affects the ease of cross-sectional deformation during bending, the radial expansion before indentation is sufficiently small compared with the inner radius of the tube and is therefore neglected in the theoretical analysis.} After the tube reaches the mechanical equilibrium, we start the indentation test. Here, the indenter is fixed throughout, while the supports are translated upward at a constant speed of 30 mm/s (sufficiently slower than the speed of the flexural wave).
The indenting force, $F$, is calculated as the sum of the components of the contact force vectors $\bm{f}_i$ at the Eulerian nodes in the downward direction $\bm{e}_{\text{down}}$, where $\bm{f}_i$ is defined in Eq.~(\ref{eq : f_i}), i.e., $F=\sum_{i\in\mathcal{G}} \bm{f}_i\cdot\bm{e}_{\text{down}}$.
We integrate the equation of motion in time up to the point where the indenter displacement reaches $\Delta = 50$~mm.

\subsection{Experimental setup}\label{sec:TubeExp}

The three-point bending test of the elastomeric tube is performed in a fish tank to minimize the effects of gravity. The elastic tube is fabricated via the dip-coating method, in which the liquid silicone elastomer coats the surface of the rigid cylinder, which is withdrawn from the polymeric bath, as the elastomer cures. The elastic tube, filled with water and pressurized from the inside, is placed underwater and bridges two separate pillars. The rigid bar connected to the load cell is lowered at a constant speed to measure the force-displacement curve (Fig.~\ref{fig:tube_result}(a)).

\subsubsection{Fabrication Protocol}

We apply the Landau-Levich coating method to polyvinyl siloxane (VPS) polymer (Elite Double 22, Zhermack, Italy) twice to fabricate a thin, homogeneous elastomeric tube. The base and catalyst solutions are mixed in equal weight (40 g total) to prepare the polymeric liquid bath. The stainless steel cylinder, 13~mm in diameter and 300~mm long, covered with the release agent (Ease Release 205, Smooth-On), is first withdrawn from the bath at constant speeds {(100~mm/min, 200~mm/min or 300~mm/min)}. The liquid polymer cures on the surface of the cylinder $\sim$20 minutes, while the coating thickness forms a gradient along the cylinder (thicker at the bottom). We turn the coated cylinder upside down and withdraw it from the newly mixed solution to form the second elastomer layer. This two-step coating protocol allows us to reduce the thickness gradient and fabricate tubes with thicknesses {$h = 1.2$~mm, 1.4~mm or 1.5~mm}. 

\subsubsection{Measurement Protocol}

The two ends of the tube are branched to connect to each (independent) water reservoir and pressure sensor (Pressure Unit~L, Fluigent, France). The reservoirs are pressurized by an air compressor (FLPG Plus, Fluigent, France). The pressure pump (FlowEZ, Fluigent, France) delivers pressurized air to extrude water from the reservoir into a tube. The independent pressure sensors at the tube extremities ensure that the tube is pressurized as homogeneously as possible along its length. {The internal pressure is set to $p = 0$~kPa, 2~kPa, 5~kPa, 8~kPa or 10~kPa}. {For the experiment of $p=0$, the tube is placed on two freely rotating ball bearings, and a three-point bending test is performed in air because no complex pressurization equipment is required. The distance between the two ball bearings is $80$~mm.} 

The load cell (LTS-2KA, Kyowa-Dengyo, Japan) is connected to the custom indenter (3~mm thick). The position of the load cell is controlled by the motorized linear stage (LTS300C/M, Thorlabs). The indenter is lowered at 1~mm/s by a distance of 50~mm. The force data are recorded by the software (DCS-100A, Kyowa-Dengyo, Japan).
We independently measure the tensile force from the adapters (inlet and outlet of the water) as $f = 0.22\tanh(0.05\Delta)$~[N]. {The tensile force from the adapters, $f$, is fitted using a hyperbolic tangent function of the indenter displacement, $\Delta$. The coefficients in the expression are determined through fitting to the experimentally measured force-displacement data when the tube elasticity is negligible.} To compare with the simulation (where the adapters are absent), we subtract $f$ from the load cell output to obtain the bending force, $F$.
We fabricate three experimental samples for each thickness and perform indentation tests 5 times per sample to obtain the experimental error bars.

\subsection{Comparison between simulation and experiment}\label{sec:TubeComparison}
\begin{figure}[h]
    \centering
    \includegraphics[width=\textwidth]{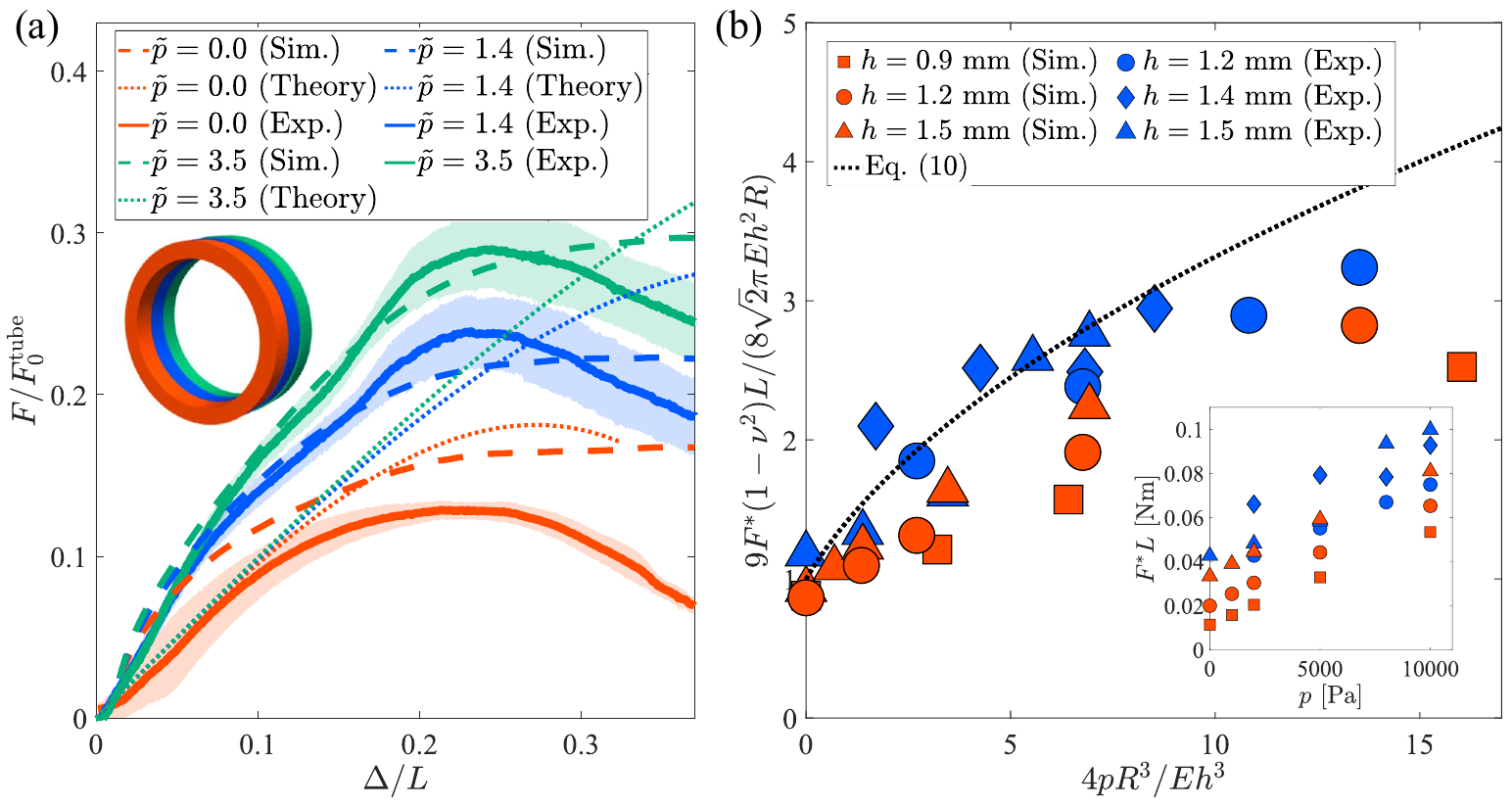}
    \caption{Comparison between experiments, simulations, and theory for the bending force of tubes. 
    (a) Comparison of the rescaled force-displacement curves for {$h = 1.5$~mm} thickness. The colored solid and dashed lines represent the experimental and numerical results, respectively. The shaded regions represent the standard deviation of the experimental results. The corresponding dotted curves represent the predictions of Eq.~\eqref{eq:FD_res}. {The insets represent the cross-section of the tube with $\Delta = 0$ for each $\tilde{p}$. The colors indicate the corresponding $\tilde{p}$.}
    (b) Comparison of the maximum force of the force displacement curve, $F^*$, for various internal pressures, $p$. Red and blue data points represent the simulation and experimental results, respectively. The same symbols represent the same thickness, $h$. The black dashed line represents Eq.~\eqref{eq:Fstar_res_p}. {The inset shows the corresponding maximum moment in dimensional form, $F^*L$, as a function of the internal pressure, $p$, before rescaling.}}
    \label{fig:tube_result}
\end{figure}

We compare the results of the three-point bending tests of the pressurized tube obtained from experiments and simulations.
The normalized force-displacement curve is shown in Fig. \ref{fig:tube_result}(a).
The force-displacement curve based on Brazier's analysis, Eq.~\eqref{eq:FD} (shown as colored dotted curves), predicts that the force increases linearly through bending, then exhibits a nonlinear response, reaching a maximum before decreasing. 
Both experimental and numerical force curves initially increase linearly and then deviate from linearity, consistent with Equation~(\ref{eq:FD}). Although the linear-to-nonlinear transitions are qualitatively consistent with theory, the force curves differ: the experimental force reaches a maximum, whereas the numerical one saturates at the critical value. {In addition, the experimental and numerical force curves reach their maximum values at $\Delta/L = 0.2$, whereas the theoretical force curve reaches its maximum at a significantly larger displacement.}
This difference would arise from three possible reasons. First, Brazier's analysis assumes that the tube is infinitely long and that its cross-sectional shape is uniform along its length, which may not be consistent with localized kink formation. Second, the extremities of the pressurized tube connect to the adapters, thereby affecting the tube’s boundary conditions, which are accounted for by applying axial forces in the simulations. Third, the fluid dynamics are not implemented in simulations. The experimental tube contains water, and the working fluid would be ejected against bending, whereas the tubes in the simulation are uniformly pressurized by the surface force. {Indeed, for $p = 0$, the theoretical maximum force is close to that obtained in the simulation}, indicating that the dynamics of the fluid inside might be crucial in this experimental setup. Implementing fluid-structure interaction is challenging in our current formalism and is left for future work.

The experimental and numerical force curves exhibit a transition from linear to nonlinear behavior, from which we compare the maximum bending force, $F^*$, across experiments, simulations, and theory. 
Based on Eq.~\eqref{eq:Fstar_res_p}, we plot the maximum bending force, $F^*$, against the internal pressure, $p$, by normalizing their characteristic values, $F_0 ^*$ and $p^*$, respectively. 
We summarize the normalized maximum bending force in experiments and simulations for various sets of the internal pressure $p$ and tube thickness $h$ in Fig.\ref{fig:tube_result}(b), comparing with Brazier's prediction Eq.~\eqref{eq:Fstar_res_p}. The maximum bending force increases with $p$, consistent with the theory, whereas the theoretical formula needs to be improved to make more quantitative predictions.

We have performed a systematic comparison of experiments, simulations, and theory for tube bending, finding slight discrepancies among them, particularly until the force reaches its peak. The difference between simulations and experiments stems from the complexity of experimental setups required to achieve sufficient precision. In the next section, we apply our formalism to the tape springs, where experimental complexities are minimized.

\section{Example 2: Open cylindrical shells}\label{sec:Tape}
We now simulate the three-point bending test to quantify the bending performance of
the tape spring. We summarize the simulation and experimental setup in Secs. ~\ref{sec:TapeSim}
and ~\ref{sec:TapeExp}, respectively. Their results are summarized and validated in Sec. ~\ref{sec:TapeComparison}.

\subsection{Simulation setting}\label{sec:TapeSim}
The three-point bending simulation of the tape spring (open cylindrical shells) is also performed using the {hybrid–MPM} framework presented in Section~\ref{sec:sim_framework}.
The tape spring, indenter, and supports are meshed as tubes within a second-order hexahedral mesh. 
The geometries of each component are defined as follows:~{The tape spring is an open cylindrical shell} with an inner radius of $R$, a thickness $h$, a subtended angle $2\beta$, and a length of 100 mm. 
The Young modulus and Poisson’s ratio are {$E = 3.8$}~GPa and {$\nu = 0.35$}, respectively. 
The indenter is a narrow rectangular plate with a height of 30 mm, a width of 40 mm, and a thickness of 1 mm, so that the deformation of the indenter is negligible.
The supports are modeled as cubes with a side length of 3 mm to ensure that the tape spring does not twist or deform out of plane. 
{The indenter is placed in the middle of the supports and the tape spring, consistent with the experimental setup (detailed next).}  The distance between the supports is $L = 80$~mm.

The simulation and measurement protocol is as follows. 
{The supports are translated upward, as in the simulation of the tubes,} at a constant speed of 100 mm/s (sufficiently slower than the flexural wave speed).
{The indenting force, $F$, is measured from the momentum transfer between the lower side of the indenter and the tape spring. The computed contact force becomes stable in time for the fixed indenter.} We integrate the equation of motion in time up to the point where the indenter displacement reaches $\Delta = 8$~mm.

\subsection{Experimental setup}\label{sec:TapeExp}
The three-point bending test of the tape spring is performed as follows.
The tape spring is fabricated thermoplastically: a plastic sheet is fitted into a cylindrical acrylic mold, heated, and then cooled to form the tape spring~\cite{Matsumoto2018, Yoshida2020, Sano2023, Yoshida2024, Matsumoto2026}.
The tape spring is placed on two rigid ball bearings, which are covered with rubber to prevent slipping.
The rigid bar connected to the load cell is lowered at a constant speed to measure the force-displacement curve (Fig.~\ref{fig:tape_result}(a)).

\subsubsection{Fabrication Protocol}
The tape springs are fabricated thermoplastically as follows. 
We laser-cut a flat rectangular piece from a polyester board of thickness {$h=0.076$~mm or $0.1$~mm} (Shim Stock, The Artus Corporation, USA) of width $w = 2\beta R$ and length 100~mm. 
The mold for the tape spring consists of two parts: open acrylic tubes and a stainless steel cylinder. The acrylic tube with an inner radius of {$R_{\mathrm{m1}} = 3.5$~mm or $7$~mm} is cut along the central axis so that the cut sheet can slide easily into the tube. We sandwiched the sheet between the open acrylic tubes and the stainless steel cylinder of radius {$R_{\mathrm{m2}} = 3.5$~mm, or $7$~mm}, and secured it with rubber bands. 
The bent sheet and mold are placed in boiling water for 2 minutes, then cooled at room temperature for 5 minutes. 
Through this process, we can fabricate a tape spring with a controlled radius of curvature $R$ and subtended angle $2\beta$. Note that this experimental protocol is used in the previous works as Refs~\cite{Matsumoto2018, Yoshida2020, Sano2023, Yoshida2024, Matsumoto2026}.

\subsubsection{Measurement Protocol}
The tape spring is mounted on two rigid ball bearings with a diameter of $D_{\mathrm{s}} = 25$~mm and a separation of $L = 80$ mm. {The ball bearings provide pin-supported boundary conditions.} The indenter is a rectangular acrylic plate (1 mm thick) and is sufficiently rigid compared with the bending of tape springs. The indenter is clamped to the force testing machine (EZ-LX, Shimadzu, Japan) and then lowered at a constant speed of 0.2 mm/s toward the center of the tape spring. We fabricate three experimental samples for each geometry, $(h,R,w) = (0.1, 6.25, 10.6), (0.076, 9.02, 15.0), (0.1, 8.85, 15.0), (0.1, 9.57, 18.5)$~mm, and perform indentation tests to obtain the experimental error bars.

\begin{figure}[!h]
    \centering
    \includegraphics[width=\textwidth]{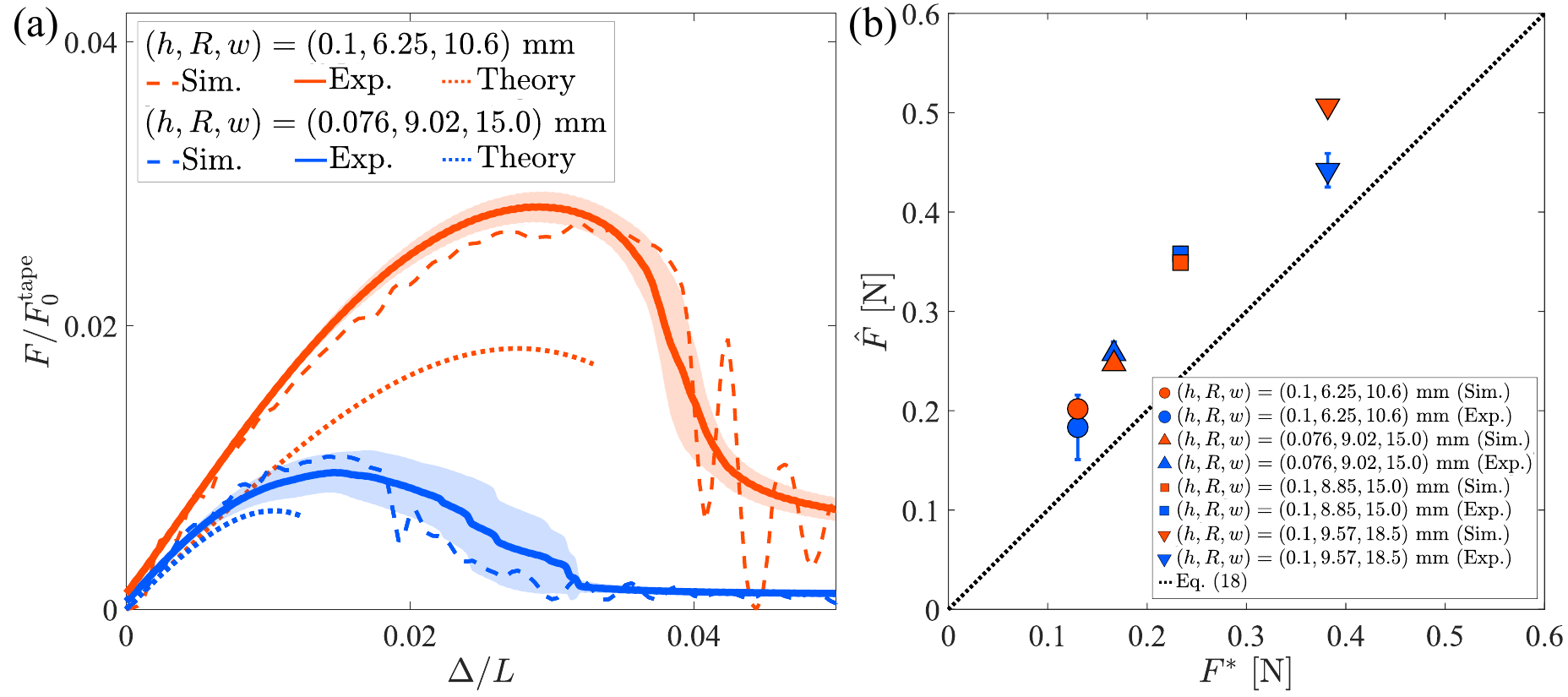}
    \caption{Comparison between experiments, simulations, and theory for tape springs. 
    (a) Comparison of the rescaled force-displacement curves. The solid and dashed lines represent the experimental and numerical results, respectively. The shaded regions represent the standard deviation of the experimental results.
    (b) Comparison of the maximum force of the force-displacement curve, $F^*$. {The horizontal axis represents the theoretical maximum force, $F^{\ast}$, Eq.~\eqref{eq:Fstar_tape}, calculated from the set of relevant parameters for each experimental and simulation setting, and the vertical axis represents the corresponding maximum force, $\hat{F}$, obtained from experiments and numerical simulations.} Red and blue data points represent the simulation and experimental results, respectively. }
    \label{fig:tape_result}
\end{figure}

\subsection{Comparison between simulation and experiment}\label{sec:TapeComparison}

We compare the experimental and simulation results for the three-point bending tests of the tape spring. The force-displacement curve is normalized based on Eq.~\eqref{eq:FD_tape_res} and summarized in Fig. \ref{fig:tape_result}(a). {The horizontal axis is limited to $\Delta/L = 0.05$ to highlight the detailed force response prior to snap-through, which occurs at approximately $\Delta = 3$~mm.} {Force responses in simulations oscillate due to transient vibrations followed by snap-through, whereas in experiments they are less pronounced. This difference would originate from differences in indentation speed or constitutive modeling. Predicting the snap-through relaxation time, as well as the agreement between experiments and simulations, would be relevant for engineering applications; we leave these as future work.}
The theoretical model based on Brazier's analysis predicts a similar linear-to-nonlinear transition in the force-displacement curve (See Eqs.~\eqref{eq:FD_res} and \eqref{eq:FD_tape_res}), consistent with our experimental and numerical results. However, the theory underestimates the force response, because the theory assumes the infinitely-long tape, thereby neglecting the finite-length effect~\cite{Seffen1999}. 
Despite the theory underestimating the measurements, the experimental and numerical results are in excellent agreement, indicating that the hybrid-MPM formalism provides precise modeling of nonlinear bending deformation in tape springs. {Here, we denote the maximum bending force obtained from the experiments and simulations by $\hat{F}$, to distinguish it from the theoretical maximum force, $F^*$, predicted by Eq. (18). Their agreement is also excellent for $\hat{F}$, shown in Fig. \ref{fig:tape_result}(b), although the theoretical prediction, $F^*$, systematically underestimates the observations.}

\section{Summary}

We have studied the bending deformation of tubes and tape springs, combining experiments, simulations, and theory. We construct experimental systems for three-point bending tests on pressurized tubes and tapes, as well as a computational framework. Our computational framework relies on the finite element method, supplemented by the material point method, to account for complex contact mechanics and replicate the experimental configurations. 

{The simulated force-displacement curve for the pressurized tube is in good quantitative agreement with the experimental results up to the displacement at which the maximum force is reached, $\tilde{\Delta}^{\ast}$.} The experimental force curve gradually decreases for $\tilde{\Delta}>\tilde{\Delta}^*$, while the simulation results saturate at $\sim\tilde{F}^*$. {This discrepancy is attributed to the difference in the boundary conditions between the experiments and simulations. In the experiments, the pressurized tube is connected to pressure adapters at both ends, whereas in the simulations the effect of the adapters is approximated by equivalent axial forces applied at the tube ends.} Despite the fact that post-buckled behaviors between experiments and simulations differ, the maximum force, $\tilde{F}^*$, is not only consistent with each other but also agrees with the classical analytical predictions. The experimental and numerical results for tape springs are in good agreement with each other, even above the kink instability. This is because the tape springs are free of end constraints, and the simulation setup more closely replicates the experimental setup. In summary, our computational framework provides the fundamental formalism to predict the bending behavior of highly flexible structures, consistent with our previous work on soft jumping robots~\cite{Abe2025}. 

{
A remaining limitation is that the proposed framework relies on an explicit dynamic formulation, which can introduce oscillations in the force response due to structural inertia, as shown in Fig.~\ref{fig:tape_result}(a). Future work will extend this hybrid-MPM framework toward quasi-static contact analysis, enabling simulations at experimental indenting speeds while retaining its robustness for large deformation and complex contact.
}

\backmatter

\bmhead{{Acknowledgements}} This work was supported by MEXT KAKENHI 24H00299 (T.G.S.), JST FOREST Program, Grant Number JPMJFR212W (T.G.S.). 

\bmhead{Competing Interests} The authors declare no competing interests.

\bmhead{{Data Availability}} {Experimental and simulation data are available upon request.}

\bmhead{Code availability}
The codes for our numerical simulations are available upon request. 

\bmhead{Authors' contributions} 
S.N., S.S., I.H., R.T., and T.G.S. designed the research and interpreted the results. S.N. and S.S. performed experiments for tape springs and tubes, respectively. S.N. implemented simulations for both tape springs and tubes. I.H. and R.T. supervised the numerical investigations. T.G.S. managed the project. 
S.N., I.H., R.T., and T.G.S. wrote the paper.


\bibliography{sn-bibliography}

\end{document}